\documentstyle[12pt,epsfig]{article}
\def\Journal#1#2#3#4{{#1}{\bf #2}, #3 (#4)}

\def\NPB{{\em Nucl. Phys.} B}

\def\PLB{{\em Phys. Lett.} B}
\def\PRL{\em Phys. Rev. Lett. }
\def\PRD{{\em Phys. Rev.} D}

\def\RMP{\em Rev. Mod. Phys. }

\begin{document}
\begin{center}
{\Large   Cosmological nucleosynthesis and active-sterile neutrino
\vspace{4mm} 
oscillations with small mass differences:\\
\vspace{0.5mm}
The resonant case}\\
\ \\
D. P. Kirilova\footnote{Permanent address: Institute of Astronomy,
Bulgarian Academy of Sciences,\\
blvd. Tsarigradsko Shosse 72, 1784 Sofia, Bulgaria}
and M. V. Chizhov\footnote{Permanent address: Centre for Space Research
and Technologies,
Faculty of Physics,\\ University of Sofia, 1164 Sofia, Bulgaria}\\
\ \\
{\it The Abdus Salam International Centre for Theoretical Physics,\\
Strada Costiera 11, 34014 Trieste, Italy}
\end{center}

\begin{flushright}
To Emmanuil, Vassillen and Rosanna
\end{flushright}

\begin{abstract}

 We have provided a numerical study of the influence of the 
resonant active-sterile neutrino oscillations 
$\nu_e \leftrightarrow \nu_s$, on the primordial production of 
helium-4. The evolution of the neutrino ensembles was followed  
selfconsistently with the evolution of the nucleons, using
exact kinetic equations for the neutrino density matrix 
and the nucleon number densities in momentum space, 
from the time of neutrino decoupling till 
the freeze-out of nucleons at 0.3 MeV.

The exact kinetic approach enabled us to study precisely  
the neutrino depletion, spectrum distortion and 
neutrino mixing generated asymmetry due to oscillations 
at each momentum mode, and to prove that their effect on 
nucleosynthesis is considerable. 

We have calculated the dependence of the primordially produced 
helium-4 on the oscillation parameters $Y_p(\delta m^2,\vartheta)$ 
for the full range of 
mixing parameters of the model of oscillations with small 
mass differences $\delta m^2 \le 10^{-7}$ eV$^2$. 
We have obtained iso-helium contours on the $\delta m^2-\vartheta$ 
plane. Cosmological constraints on oscillation 
parameters, more precise   
than the existing ones were extracted, due to the 
exact kinetic approach and the proper account for the 
neutrino spectrum distortion and the oscillations generated 
asymmetry. 
 
\end{abstract}

\section{Overview of neutrino oscillations and primordial nucleosynthesis}

   Nowadays several major experiments point to the existence of 
neutrino oscillations. Besides,
the neutrino puzzles, namely, the solar neutrino deficit and the 
atmospheric neutrino anomaly, which are believed to be 
explained in terms of neutrino oscillations,  are still 
 with us. 
Therefore, it seems useful  to obtain  more precise 
cosmological constraints on neutrino oscillation parameters. 
Moreover, the very low values of mass differences, 
explored in the cosmological model we will discuss, 
are  beyond the reach of 
present and near future experimental constraints. 
In the present work we explore the effect of resonant active-sterile 
neutrino oscillations with small mass
differences $\delta m^2 \le 10^{-7}$ eV$^2$ 
on the primordial production of helium-4 and obtain precise 
cosmological constraints on the neutrino oscillation parameters.  

The problem of neutrino oscillations and Big Bang Nucleosynthesis (BBN) 
has been discussed in numerous 
publications~\cite{D81}-\cite{new}. 
First cosmological constraints from       
BBN on oscillation parameters were obtained analytically in~\cite{BD}.
Then numerical calculations of the oscillations effect on primordial 
nucleosynthesis were made  in~\cite{Enqvist}. In these works an account  
 for  the depletion of the neutrino number densities due to oscillations
 was provided, while  the neutrino-antineutrino asymmetry and the
distortion of the neutrino spectrum were neglected (see also \cite{ssf}). 
The importance of the neutrino spectrum distortion for the 
BBN with oscillations
was first noticed for the vacuum oscillation case in ref.~\cite{K88}.
First precise account of both neutrino spectrum distortion and the 
oscillations generated neutrino-antineutrino asymmetry  effects in  BBN
calculations and cosmological 
constraints on oscillation parameters were provided in ref.~\cite{NU96}.   

   The problem of active-sterile {\it non-resonant} neutrino
oscillations and the
primordial helium-4 production was thoroughly
investigated and completed  
for the model of nonequilibrium oscillations with small mass differences 
~\cite{K88,NU96} in refs.~\cite{PR,NP}. There we have analyzed 
numerically the role of oscillations in BBN
using the exact kinetic equations for the neutrino density matrix and 
nucleon number densities in momentum space. The exact kinetic approach enabled 
us to reveal the important role of the spectrum distortion and neutrino 
population depletion. 
On the other hand, for the nonresonant case it was shown 
that the lepton asymmetry can be neglected in case initially
it was of the order of the baryon one~\cite{PR}, while in case it was 
greater than 10$^{-7}$ it 
may considerably influence BBN~\cite{NP}. 
Precise constraints on the 
oscillation parameters by almost an order of magnitude better 
than the existing ones~\cite{Enqvist,ssf} concerning the neutrino squared mass
differences,
were obtained due to the exact kinetic approach and selfconsistent 
account of the evolution of the neutrinos  and the nucleons.
The analytical fit to the exact constraints~\footnote{ The constraints are 
derived for primordial helium-4 $Y_p \le 0.24$. Other iso-helium contours 
and the corresponding constraints were calculated in refs.~\cite{PR,NP}} is: 
\begin{equation}
\delta m^2 (\sin^2 2\vartheta)^4 \le 1.5\times 10^{-9}~{\rm eV}^2
\end{equation} 
for $\delta m^2 \le 10^{-7}$ eV$^2$.

It was noticed in ref.~\cite{NU96}, that
the {\it resonant} neutrino oscillations case is a much more complicated
one, as far
as rapid growth of asymmetry for certain sets of parameters is
typical there~\cite{BD,FTV96,Shi96,NU96}. 
First detailed calculations of the
BBN with resonant neutrino oscillations accounting for the
asymmetry growth were provided in~\cite{NU96}. 
The phenomenon of the oscillation-generated asymmetry growth was 
registered there and the calculations of the BBN were provided accounting
both for the
spectrum distortion and for the neutrino asymmetry dynamical evolution
{\it at each momentum mode}.
It was shown that following the behavior of the neutrino-antineutrino asymmetry 
 at each momentum is important, particularly, when the distortion of 
the neutrino spectrum is considerable. 
As a result helium-4  contour $Y_p=0.245$ was obtained and precise
constraints on the
oscillation parameters were provided. They are  better by almost an order of
magnitude
than the existing ones for the neutrino squared mass differences 
at large  mixing angles.  

In ref.~\cite{new} the effect of the neutrino-mixing-generated
asymmetry was shown to be considerable --
up to about $10\%$ relative decrease in helium-4 in comparison with the 
case with oscillations but without the asymmetry account.
Hence, a more profound study of the BBN with resonant oscillations, accounting
properly for the important asymmetry effect  is necessary. 
The purpose of the present work is to provide a more detail study of 
the BBN with resonant neutrino oscillations.

\section{Nucleosynthesis with oscillating neutrinos}

In the present  work we expand  and complete the original 
investigations~\cite{NU96,new} of the asymmetry
effect
on primordial production of helium in the model of BBN with resonant
neutrino oscillations, for the full parameter space of the 
nonequilibrium oscillations model~\cite{NU96}.

We consider the case of active-sterile neutrino
oscillations with small mass differences, 
namely $\delta m^2 \le 10^{-7}$ eV$^2$, 
described in detail elsewhere~\cite{NU96,PR}, where
the nonequilibrium effects are stronger and, therefore, 
it is less studied one till now. According to that model, 
oscillations proceed effectively after the active neutrino decoupling and
till
then the sterile neutrinos have not yet thermalized, so that their number
density 
is negligible in comparison with the electron neutrino one.
For simplicity we assume mixing just in the electron sector,
$\nu_i=U_{il}~\nu_l$ ($l=e,s$).

The set of kinetic equations describing simultaneously
the evolution of the  neutrino and antineutrino density matrix
$\rho$ and  $\bar{\rho}$  and the   
evolution of the neutron
number density   $n_n$ in momentum space reads:

\begin{eqnarray}
&&{\partial \rho(t) \over \partial t} =
H p_\nu~ {\partial \rho(t) \over \partial p_\nu} +
\nonumber\\
&&+ i \left[ {\cal H}_o, \rho(t) \right]
+i \sqrt{2} G_F \left(\pm {\cal L} - Q/M_W^2 \right)N_\gamma
\left[ \alpha, \rho(t) \right]
+ {\rm O}\left(G_F^2 \right),
\label{kin}
\end{eqnarray}
\begin{eqnarray}
&&\left(\partial n_n / \partial t \right)
 = H p_n~ \left(\partial n_n / \partial p_n \right) +
\nonumber\\
&& + \int {\rm d}\Omega(e^-,p,\nu) |{\cal A}(e^- p\to\nu n)|^2
\left[n_{e^-} n_p (1-\rho_{LL}) - n_n \rho_{LL} (1-n_{e^-})\right]
\nonumber\\
&& - \int {\rm d}\Omega(e^+,p,\tilde{\nu}) |{\cal A}(e^+n\to
p\tilde{\nu})|^2
\left[n_{e^+} n_n (1-\bar{\rho}_{LL}) - n_p \bar{\rho}_{LL} (1-n_{e^+})\right].
\end{eqnarray}
where $\alpha_{ij}=U^*_{ie} U_{je}$,
$p_\nu$ is the momentum of electron neutrino,
 $n$ stands for the number density of the interacting particles,
${\rm d}\Omega(i,j,k)$ is a phase space factor and  ${\cal A}$ is the
amplitude of the corresponding process.
The sign plus in front of ${\cal L}$ corresponds to neutrino
ensemble, while minus - to antineutrino ensemble.

The initial condition for the
neutrino ensembles in the interaction basis
is assumed of the form:
$$
{\cal \rho} = n_{\nu}^{eq}
\left( \begin{array}{cc}
1 & 0 \\
0 & 0
\end{array} \right)
$$
where $n_{\nu}^{eq}=\exp(-E_{\nu}/T)/(1+\exp(-E_{\nu}/T))$.

It corresponds to the standard equilibrium distribution of
active electron neutrinos, and an absence of the sterile ones.
The initial values for the neutron, proton and electron
number densities are their equilibrium values.

The first term in the right hand side of the equations (2)
and (3)  describes the effect of
Universe expansion.
The second term in (2) is responsible for neutrino oscillations, the
third accounts for forward neutrino scattering off the
medium and the last one accounts for the second order interaction
effects of neutrinos with the medium.
It is important for the nonequilibrium active-sterile neutrino oscillations
to provide simultaneous account of the different competing processes,
namely: neutrino oscillations, Hubble expansion and weak interaction processes.
${\cal H}_o$ is the free neutrino Hamiltonian.
The `nonlocal' term $Q$ arises as an $W/Z$ propagator effect,
$Q \sim E_\nu~T$.
${\cal L}$ is proportional to the fermion asymmetry of the plasma
and is essentially expressed through the neutrino asymmetries
${\cal L} \sim 2L_{\nu_e}+L_{\nu_\mu}+L_{\nu_\tau}$,
where
$L_{\mu,\tau} \sim (N_{\mu,\tau}-N_{\bar{\mu},\bar{\tau}})/ N_\gamma$
and $L_{\nu_e} \sim \int {\rm d}^3p (\rho_{LL}-\bar{\rho}_{LL})/N_\gamma$.

The neutron and proton number
densities, used in the kinetic equations for neutrinos eq.~(2),
were substituted
from the numerical calculations of eq.~(3). On the other hand, 
$\rho_{LL}$ and $\bar{\rho}_{LL}$
at each integration step of eq.~(3) was taken from the simultaneously
performed integration of the set of equations (2). 
I.e.\ we have selfconsistently followed 
 the evolution of neutrino ensembles and the nucleons.

 We account for the exact
kinetics both of the neutrino and the neutron-proton transition, essential
for the
helium-4 synthesis. 
Besides, the equations  follow neutrino evolution
in momentum space, i.e.\ enabling to account 
 accurately for the neutrino
depletion,
neutrino energy spectrum distortion and the dynamical
evolution of the asymmetry.

Eq.~(2) results into a set of coupled nonlinear
integro-differential equations with time dependent coefficients
for the components
of the density matrix of neutrinos:
four  equations for the components of
the neutrino density matrix, and  another four  for
 the  antineutrino density matrix  for each momentum mode.
However,
due to conservation of the total neutrino number density
in the discussed model, the number of the equations 
can be reduced to 6 equations for each momentum mode
of neutrinos and antineutrinos.

The equations were integrated for the
characteristic period from the electron neutrino decoupling
at 2 MeV till the $n/p$ freeze-out at 0.3 MeV.
We have calculated  the yields of 
primordially produced helium-4  for  
the full range of the model's parameters values, namely
for $\sin^2(2\vartheta)$ ranging from $10^{-3}$ to maximal mixing
and $10^{-11}$ eV$^2 \le \delta m^2 \le 10^{-7}$ eV$^2$.
For smaller mixing parameters the effect on
helium-4 was shown to be  negligible~\cite{NU96}.

Our results are based on hundreds of $\delta m^2-\vartheta$ combinations. 
The spectrum distribution we have usually described by 1000 bins. 
Mind, however, that 
for some sets of parameters, where rapid growth of asymmetry occurs,  
even 5000
bins do not give satisfactory good description of the great spectrum 
distortion 
and the rapid sign changing behavior of the asymmetry.
Fortunately, we have estimated the effect of this numerical uncertainty on the
calculated production of helium-4 and found that it is much 
less than $1\%$ for the full oscillation parameters range. 

Therefore, we are not discussing here the asymmetry behavior, but present 
only the results of our study concerning nucleosynthesis 
 which are trustable.
The analysis of the precise  asymmetry evolution itself deserves further 
investigation. We are quite convinced by our studies, that, surely, the
real physical
behavior of the asymmetry should not be a function of the calculational 
parameters,
such as different error control, step size, et cetera.
(See, however, the opposite point of view on that question by Shi 
in \cite{Shi96}).
According to us, such kind of a dependence on the calculational parameters 
points only to the unsatisfactory accuracy  of the numerical calculations
 or of the calculational methods used.  The
question is even more complicated, as far as we have 
estimated that  the neutrino evolution 
equations at resonance have  high stiffness. 
Hence, the usual explicit numerical approach is not applicable
for the description of the asymmetry evolution,
especially, if the correct account for the spectrum spread of neutrino is 
provided.
To solve the stiff equations numerically, implicit methods should be used. 
For 1000 bins of the spectrum  a system of 6000 equations describing the
neutrino density
evolution should be solved simultaneously. And this is a hopeless task
with our facilities now. 
We will  discuss this question in more detail elsewhere. 
 
\section{Results and conclusions}

The major effects, of the discussed resonant 
$\nu_e \leftrightarrow \nu_s$ oscillations with small 
mass differences on helium-4 production, are due to 
the depletion of the neutrino number densities, 
neutrino spectrum distortion and the neutrino asymmetry 
growth due to oscillations.   

(a) Depletion of $\nu_e$ population due to oscillations:

 As far as oscillations become effective when the number densities
of $\nu_e$ are much greater than those of $\nu_s$,
 the oscillations tend to reestablish
the statistical equilibrium between different oscillating species.
As a result $\nu_e$ decreases in comparison to its standard
equilibrium value due to oscillations
in favor of sterile neutrinos.
The depletion of the electron neutrino number densities due to
oscillations into sterile ones strongly affects the
$n \leftrightarrow p$ reactions rates. It leads to an effective 
decrease in the weak
 processes rates, and, hence, to an increase of the freezing
temperature of the $n/p$-ratio and the corresponding overproduction of the
primordially produced $^4\! He$.

(b) Distortion of the energy distribution of neutrinos:

Neutrinos with different momenta begin to oscillate at different
temperatures and with different amplitudes.
First the low energy part of the spectrum is distorted, and later on
this distortion concerns neutrinos with higher and higher energies.
The effect of the distortion of the energy distribution of neutrinos
on helium-4 production is two-fold. On one hand an average decrease 
of the energy of active neutrinos leads to a decrease of the weak 
reactions rate, and hence, to an increase in the
freezing temperature and the produced helium. On the other hand,
there exists an energy threshold for the reaction
$\tilde{\nu}_e+p \to n+e^+$. So, in case when
the energy of the relatively greater part of neutrinos becomes
smaller than that threshold the $n/p$-freezing ratio decreases
leading to a corresponding decrease of the primordially produced
helium-4~\cite{ki}. The numerical analysis showed that  
 the total effect of the  distortion of the energy distribution 
is  an increase in the produced helium.

(c) Neutrino asymmetry:

Neutrino mixing generated asymmetry effect was found to be considerable.
(See also ref.~\cite{new}).
It was proven numerically, that in  the case of
small mass differences we
discussed and naturally small initial asymmetry,  
the growth of the asymmetry is less than 4 orders of magnitude. Hence,
beginning with asymmetries of the order of the baryon one,
the asymmetry does not grow enough
to influence {\it directly} the kinetics of the  $n-p$ transitions.
Consequently, the apparently great
asymmetry effect (as illustrated in Fig.~2)
is totally due to the {\it indirect} effects of the asymmetry
on BBN.\@
The maximal asymmetry effect is around $10\%$ `underproduction' of
$Y_p$ in comparison with the case of BBN with oscillations but
without the asymmetry account.

The total effect of oscillations, with the complete
account of the asymmetry effects, is still overproduction of
helium-4, in comparison to the standard value, 
although considerably smaller at small mixing angles
than in the calculations neglecting asymmetry. Therefore,
nucleosynthesis constraints on the mixing parameters of neutrino
are alleviated considerably due to the asymmetry effect.

From the numerical integration for the full range of
 oscillation parameters
we have obtained the primordial helium yields
$Y_p(\delta m^2,\vartheta)$.
Some of the iso- helium
contours calculated in the discussed model of cosmological
nucleosynthesis with resonant neutrino oscillations
are presented on the plane $\delta m^2-\vartheta$ in Fig.~1.

At present the primordial helium values
extracted from observations differ considerably~\cite{CN}.
Therefore, we consider
it useful to provide the exact calculations
for various iso-helium contours up to 0.26.
Knowing more precisely the primordial
helium-4 value from  observations, it will be possible to
obtain the excluded region of the mixing parameters using the results of
this survey.
For example, assuming the `low' observational value of  
primordial $^4\! He$ $Y_p \cong 0.234$~\cite{CN},
the cosmologically excluded region for the oscillation parameters
is situated on the plane $\delta m^2-\vartheta$ 
to the right of the $Y_p=0.245$ curve, which
gives $5\%$ overproduction of helium in comparison with this
observational value.

In  Fig.~2  a comparison between the curves, corresponding
to helium abundance $Y_p=0.24$, obtained in the present work and
in previous works~\cite{Enqvist,ssf}, 
analyzing the resonant active-sterile neutrino
oscillations, is presented.
In \cite{Enqvist} the excluded
regions for the neutrino mixing parameters were obtained from the
requirement that the neutrino types should be less than 3.4:
$N_{\nu}<3.4$. The depletion effect was considered, while 
the neutrino-antineutrino asymmetry was neglected, and the 
distortion of the neutrino spectrum
was not studied,  instead  the kinetic equations for neutrino mean
number densities were used.

The dashed curve, presenting our results, in case
the asymmetry effect was neglected, is in a good accordance with the
results of Enqvist et al.~\cite{Enqvist}, where asymmetry was neglected.
The difference between the two curves shows explicitly the effect of
the proper account of the neutrino spectrum spread and spectrum distortion,
which was provided in our work.
On the other hand, the difference between our curves, the solid and
the dashed one, presents the net asymmetry effect.

The results of~\cite{ssf} 
differ both from the ones of ref.~\cite{Enqvist} and from
our results. 
We consider them not correct. Our conclusion is not only based 
on the discrepancy between these results and those of other studies, 
but on the very fact that they are not consistent even between themselves
concerning resonant and nonresonant case.
As is well known from the analytical formulae the results for the resonant case 
$\delta m^2 < 0$ coincide with those for the 
nonresonant one $\delta m^2 > 0$ at maximal mixing. 
This fact is illustrated in Fig.~3 of resonant and nonresonant 
oscillations for all studies, except ref.~\cite{ssf}.

As is seen from the iso-helium contours for $Y_p=0.24$,
for {\it large mixing angles}
we exclude  mass differences $\delta m^2\ge 8.2\times 10^{-10}$ eV$^2$, 
which is an order
of magnitude stronger constraint than the previously existing.
This more stringent constraint for mass differences, 
obtained in our work for the region of
large mixing angles  is due to
the more accurate kinetic approach we have used
and to the precise
account of neutrino depletion and energy distortion.
On the other hand, at {\it small mixing angles} the account of 
the oscillations generated asymmetry leads to an alleviation of the 
BBN constraints in comparison with the previous 
works~\cite{Enqvist,ssf}. 
It is easy to understand, as far as the  asymmetry growth results in 
suppression of oscillations and, hence, less strongly pronounced 
overproduction of helium-4 due to oscillations than in the case without the
asymmetry account. 

In conclusion, we have shown that both the spectrum distortion and 
neutrino mixing generated asymmetry
should be accounted for properly in models of BBN with oscillations,
as far as their effect is considerable. We have calculated different 
iso-helium
contours for the resonant case of neutrino oscillations with small mass
differences. The cosmological constraints obtained  are better by an order of 
magnitude than the existing ones due to the exact kinetic approach both to the
neutrino evolution and to the nucleons freeze-out.\\ \ 

\section{Acknowledgements}

We highly acknowledge the hospitality and the support of ICTP, Trieste, 
during the preparation of this work. 
We are grateful to  S. Randjbar-Daemi for the 
opportunity to work at ICTP. 
D.K. thanks D. Sciama for the possibility to participate 
into the astrophysics program this summer, 
which was essential for the successful completing of this
survey.

D.K is glad to thank V. Semikoz for stimulating discussions and
encouragement at the beginning of her research 
on matter oscillations and BBN in 1994.  
We would like also to express our gratitude to    
A. Dolgov for fruitful discussions and P. Christensen
for the overall help. Part of the numerical calculations 
were provided using  computational powers of the  Theoretical 
Astrophysics Center, Copenhagen. 
This work was supported in part by the Danish National
Research Foundation through its establishment of the Theoretical
Astrophysics Center.

We are obliged also to Emmanuil, Vassillen and Rosanna for their 
great patience during our work on this theme.

\pagebreak[1]

\newpage
\pagestyle{empty}

{\bf Figure captions}\\

{\bf Figure 1.}  On the $\delta m^2-\vartheta$ plane iso-helium-4
contours $Y_p=0.24$, 0.245, 0.25, 0.255 and 0.26,
calculated in the discussed model of BBN with active-sterile
resonant neutrino oscillations are  shown.
For fixed primordial helium-4 value,
the area to the left of the corresponding curve gives
the allowed region of the oscillation parameters.\\

{\bf Figure 2.} In the figure a comparison between the results concerning 
primordial helium-4 production, obtained in the present work and  
previous works~\cite{Enqvist,ssf}, is presented. 
The dashed curve shows our results in case without
asymmetry effect account. It is in a good accordance with the 
results of Enqvist et al.~\cite{Enqvist}, 
where asymmetry was neglected. 
The difference between the two curves shows explicitly the effect of 
the proper account of the spectrum spread  of neutrino, 
which was provided in our work. 
On the other hand, the difference between our curves, the solid and 
the dashed one presents the net asymmetry effect.
The artistic curve of Shi et al.~\cite{ssf} is obviously inconsistent
with the results of other works and we will leave it without
a comment.\\

{\bf Figure 3.} Combined  iso-helium contours $Y_p=0.24$, 
for the resonant oscillations, $\delta m^2 < 0$, and the nonresonant 
ones, $\delta m^2 > 0$, calculated in previous 
studies~\cite{Enqvist,ssf,PR} and in this work, are presented.
The discontinuity of the curve of Shi et al.~\cite{ssf} reveals the 
discrepancy between their own results for the resonant and 
nonresonant case. 

\newpage
\pagestyle{empty}
\begin{figure}
\epsfig{file=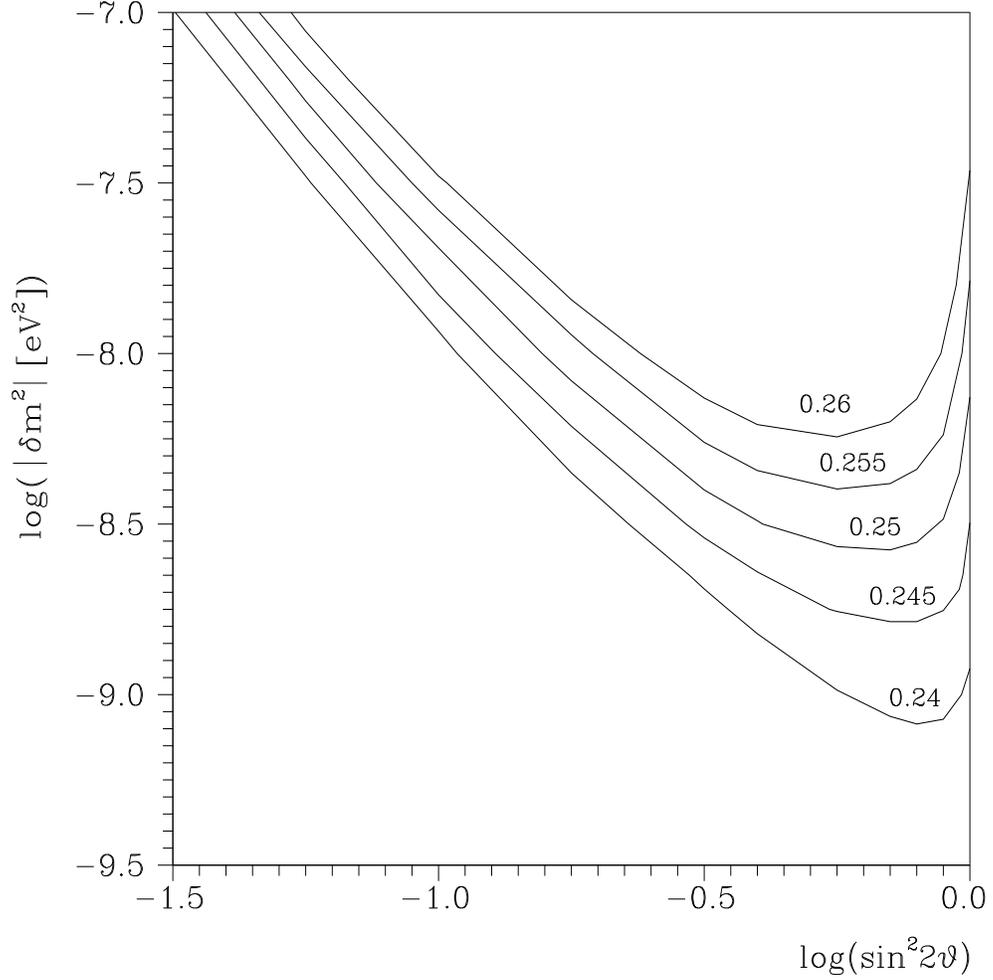,height=13cm,width=13cm}
\caption{On the $\delta m^2-\vartheta$ plane iso-helium-4
contours $Y_p=0.24$, 0.245, 0.25, 0.255 and 0.26,
calculated in the discussed model of BBN with active-sterile
resonant neutrino oscillations are  shown.
For fixed primordial helium-4 value,
the area to the left of the corresponding curve gives
the allowed region of the oscillation parameters.}
\end{figure}

\newpage
\pagestyle{empty}
\begin{figure}
\epsfig{file=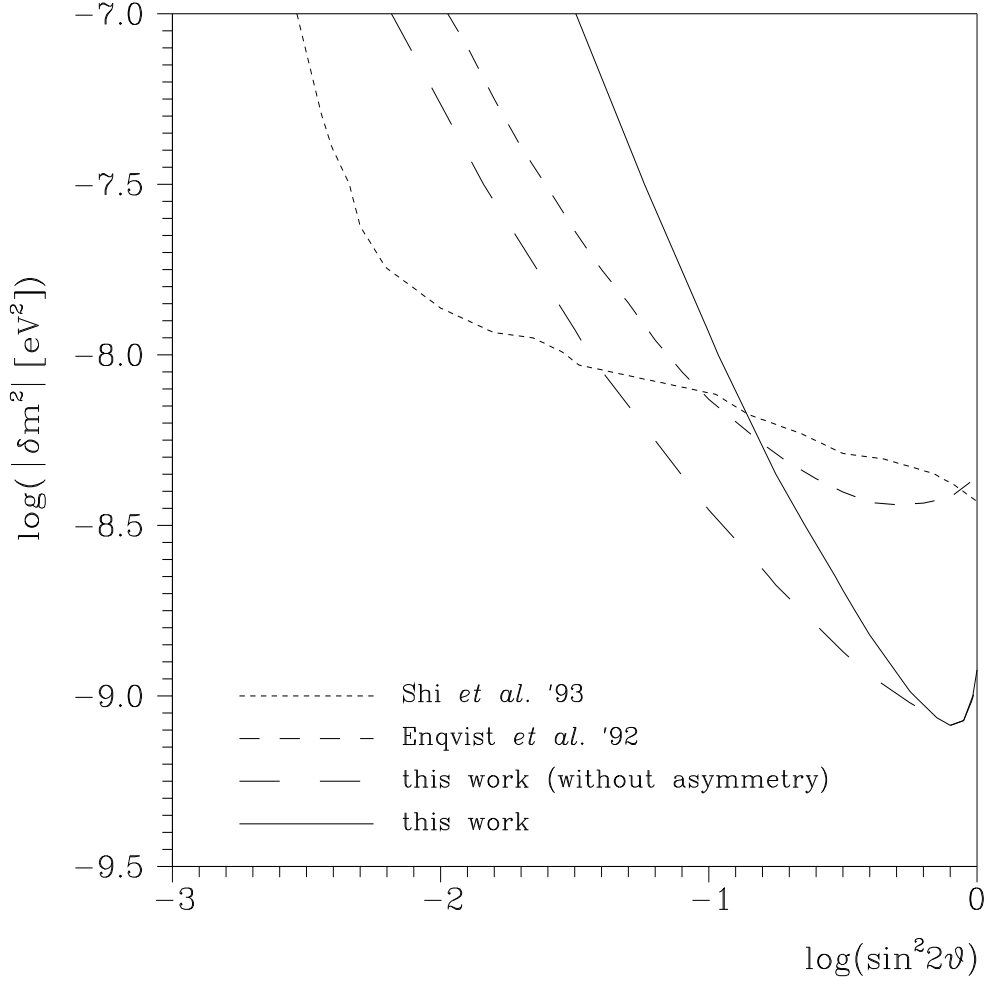,height=13cm,width=13cm}
\caption{In the figure a comparison between the results concerning
primordial helium-4 production, obtained in the present work and
previous works~\cite{Enqvist,ssf}, is presented.
The dashed curve shows our results in case without
asymmetry effect account. It is in a good accordance with the
results of Enqvist et al.~\cite{Enqvist},
where asymmetry was neglected.
The difference between the two curves shows explicitly the effect of
the proper account of the spectrum spread  of neutrino,
which was provided in our work.
On the other hand, the difference between our curves, the solid and
the dashed one presents the net asymmetry effect.
The artistic curve of Shi et al.~\cite{ssf} is obviously inconsistent
with the results of other works and we will leave it without
a comment.}
\end{figure}

\newpage
\pagestyle{empty}
\begin{figure}
\epsfig{file=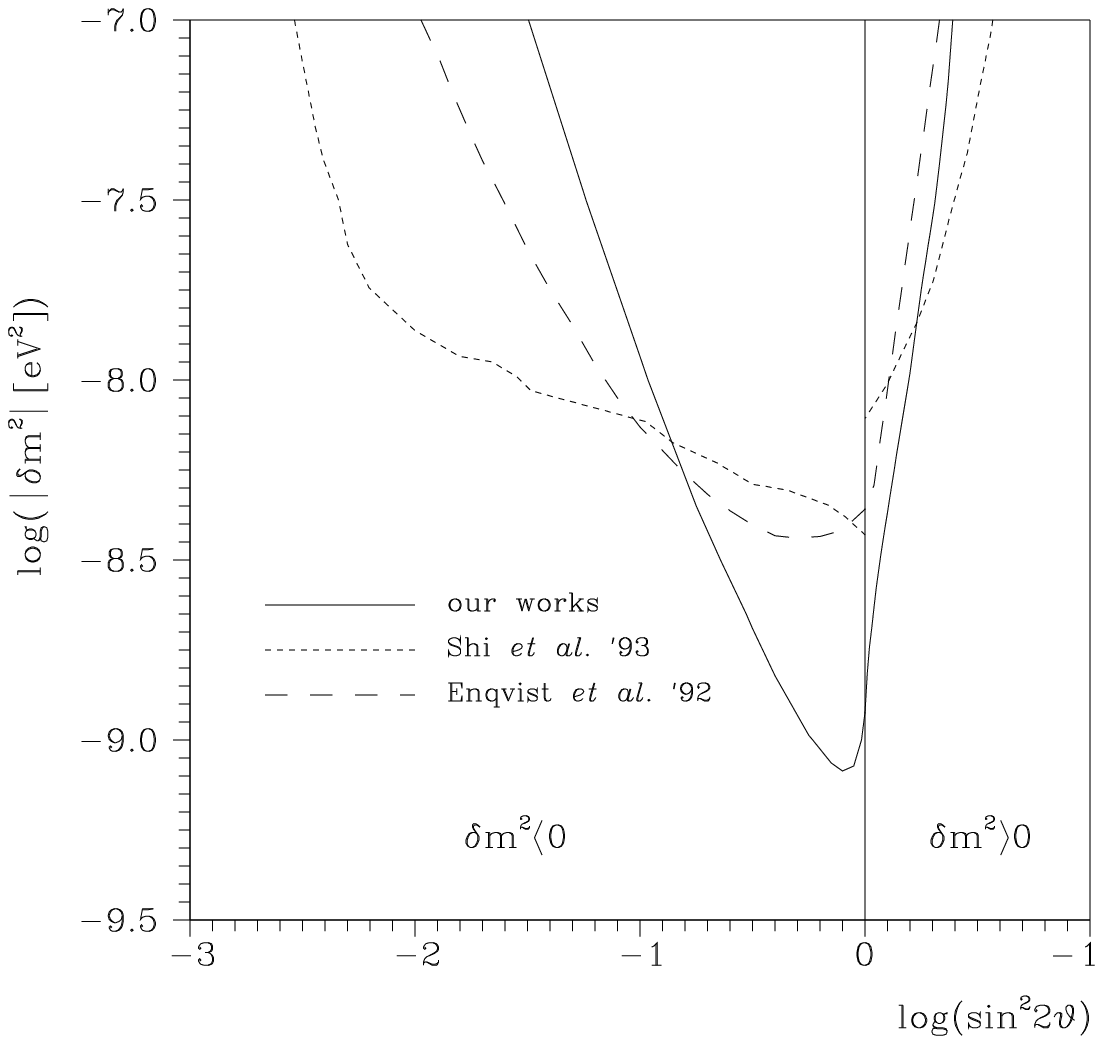,height=13cm,width=13cm}
\caption{Combined  iso-helium contours $Y_p=0.24$,
for the resonant oscillations, $\delta m^2 < 0$, and the nonresonant
ones, $\delta m^2 > 0$, calculated in previous
studies~\cite{Enqvist,ssf,PR} and in this work, are presented.
The discontinuity of the curve of Shi et al.~\cite{ssf} reveals the
discrepancy between their own results for the resonant and
nonresonant case.}
\end{figure}


\begin{thebibliography}{99}

\bibitem{D81} A. D. Dolgov, Sov. J. Nucl. Phys. 33, 700 (1981).

\bibitem{hp} M. Yu. Khlopov and S. T. Petcov,
\Journal{\PLB}{99}{117}{1981}, Erratum: \Journal{\PLB}{100}{520}{1981}.

\bibitem{ns} D. Fargion and M. Shepkin, \Journal{\PLB}{146}{46}{1984}.

\bibitem{MSW} A. Yu. Smirnov, talk given at the VI Moriond Meeting on 
 massive neutrinos in particle physics and astrophysics, Tignes, France,
 1986.

\bibitem{lan} P.G. Langacker, B. Sathiadalan, and G. Steigman, 
\Journal{\NPB}{266}{669}{1986};\\
P. Langacker, S. Petcov, G. Steigman, and S. Toshev,
\Journal{\NPB}{282}{589}{1987}.

\bibitem{K88} D. P. Kirilova, JINR E2-88-301, 1988.

\bibitem{Kuo} T. K. Kuo and J. Panteleone,
\Journal{\RMP}{61}{937}{1989}.

\bibitem{BD}R. Barbieri and A. Dolgov, Phys. Lett. B 237, 440 (1990);\\
R. Barbieri and A. Dolgov, Nucl. Phys. B 349, 743 (1990);\\
K. Kainulainen, Phys. Lett. B 244, 191 (1990).

\bibitem{savage} M. G. Savage, R. Malaney, and G. Fuller, Astrophys. J.
368(1991).

\bibitem{Enqvist} K. Enqvist, K. Kainulainen, and J. Maalampi,
Phys. Lett. B 244, 186 (1990); Nucl. Phys. B 349, 754 (1991);\\

\bibitem{cline} J. M. Cline, \Journal{\PRL}{68}{3137}{1992}.

\bibitem{sr} G. Sigl and  G. Raffelt,
\Journal{\NPB}{406}{423}{1993}.

\bibitem{ssf} X. Shi, D. N. Schramm, and B. D. Fields,
\Journal{\PRD}{48}{2563}{1993}.

\bibitem{FV95} R. Foot and R. Volkas, Phys. Rev. Lett. 75, 4350 (1995).

\bibitem{Kost} V. A. Kostelecky and S. Samuel, Phys. Lett. B 385, 159
(1996); Phys. Rev. D 52, 3184 (1995).


\bibitem{FTV96} R. Foot, M. J. Thompson, and R. R. Volkas,
Phys. Rev. D 53, R5349 (1996).

\bibitem{card} C. Y. Cardall and G. M. Fuller, 
Phys. Rev. D 54, R1260 (1996).

\bibitem{Shi96} X. Shi, Phys. Rev. D 54, 2753 (1996).

\bibitem{NU96} D. P. Kirilova and M. V. Chizhov,
in Proc. NEUTRINO 96 Conference, Helsinki, 1996, p. 478;\\
D. P. Kirilova and M. V. Chizhov, Phys. Lett. B 393, 375 (1997).  

\bibitem{FVlast}  R. Foot and R. Volkas, Phys. Rev. D 55, 5147 (1997);\\
Phys. Rev. D 56, 6653 (1997); Erratum ibid. D 59, 029901 (1999);\\
N. Bell, R. Foot, and R. Volkas, Phys. Rev. D 58, 105010 (1998);\\
R. Foot and R. Volkas, astro-ph/9811067;\\
  R. Foot and R. Volkas, hep-ph/9904336;\\
R. Foot, Astropart. Phys. 10, 253 (1999);\\
R. Foot, hep-ph/9906311.

\bibitem{PR} D. P. Kirilova and M. V. Chizhov, Phys. Rev. D 58, 073004
(1998).

\bibitem{NP} D. P. Kirilova and M. V. Chizhov, Nucl. Phys. B 534, 447
(1998).

\bibitem{wrong} K. Abazajian, X. Shi, and G. Fuller, astro-ph/9904052;\\
X. Shi, G. Fuller, and K. Abazajian, astro-ph/9908081;\\
X. Shi, G. M. Fuller, and K. Abazajian, Phys. Rev. D 59, 063006 (1999);\\
X. Shi and G. Fuller, astro-ph/9904041;\\
X. Shi, G. M. Fuller, and K.Abazajian, Phys. Rev. D 60, 063002 (1999);\\

\bibitem{new} M. V. Chizhov and D. P. Kirilova, hep-ph/9908525.

\bibitem{ki} A. Dolgov and D. Kirilova, Int. J. Mod. Phys. A3, 267 (1988).

\bibitem{CN} G. Steigman, astro-ph/9803055;\\
S. Sarkar, astro-ph/9903183;\\
K. Olive, G. Steigman, and T. Walker, astro-ph/9905320;\\
S. Esposito et al., astro-ph/9906232.

\end{thebibliography}
\end{document}